\documentclass[aps,prb,twocolumn,showpacs,superscriptaddress]{revtex4}
\usepackage{graphicx}
\usepackage{bm}

\newcommand{\beq}{\begin{equation}}
\newcommand{\eeq}{\end{equation}}

\begin{document}

\title{\boldmath Anomalous Behavior of the Spin Susceptibility
                 of Strongly Correlated Fermi Systems }

\author{J.~W.~Clark}
\affiliation{ McDonnell Center for the Space Sciences and
Department of Physics, Washington University,
St.~Louis, MO 63130, USA }
\author{V.~A.~Khodel}
\affiliation{ Russian Research Centre Kurchatov
Institute, Moscow, 123182, Russia}
\affiliation{ McDonnell Center for the Space Sciences and
Department of Physics, Washington University,
St.~Louis, MO 63130, USA }
\author{M.~V.~Zverev}
\affiliation{ Russian Research Centre Kurchatov
Institute, Moscow, 123182, Russia}

\date{\today}

\begin{abstract}

The spin susceptibility $\chi(T)$ of strongly correlated Fermi systems
is investigated in the density region where Fermi-liquid theory fails.
We attribute this failure to a specific quantum phase transition
associated with a rearrangement of the Landau state at low temperatures
$T$, retaining the assumption that the Landau quasiparticle picture
survives in a generic sense.  Taking into account the resulting
modification of the quasiparticle distribution function, the spin
susceptibility $\chi(T)$ is shown to contain a Curie-Weiss component
$\chi_{\rm CW}(T)\sim (T-\Theta_{\rm W})^{-1}$, with the Weiss
temperature $\Theta_{\rm W}$ vanishing at the critical density for
the transition.
\end{abstract}

\pacs{
71.10.Hf,
71.27.+a
}
\maketitle

The manifestation of non-Fermi-liquid behavior in strongly correlated
Fermi systems provides valuable clues to a fundamental microscopic
understanding of these systems.  This letter was stimulated by the
findings of recent studies of laboratory simulants of two-dimensional
(2D) liquid $^3$He and the electron
gas.\cite{godfrin1,godfrin2,saunders,krav,rez,okamoto,vitkalov,pudalov1}
Two dramatic effects have been revealed by data for the spin
susceptibility $\chi(T)$ and specific heat of 2D liquid $^3$He
and for the temperature dependence of Shubnikov-de Haas conductivity
oscillations in the 2D electron gas.  First, the results point
to a divergence of the effective mass $M^*$ in these systems at
a critical density $\rho_{\infty}$ where, concomitantly, the
single-particle spectrum $\varepsilon(p)$ becomes flat.  Second,
as seen in the experimental data\cite{godfrin1,godfrin2,rez}
on $\chi(T\to 0)$, Landau theory begins to fail even before
this critical density is reached.

Ordinarily, the non-Fermi-liquid behavior of $\chi(T\to 0)$ is
attributed to a disorder-driven localization transition.  To be sure,
disorder plays a part, as evidenced in data\cite{sheldon} on the
spin susceptibility of submonolayer $^3$He at density
$\rho=0.0064$\,\AA$^{-2}$($\sim 0.1$ layer), absorbed onto a
Nuclepore substrate pre-plated with several $^4$He layers. The
spin susceptibility of this low-density disordered system is
found to contain a significant Curie component $\chi_C(T)\sim T^{-1}$,
which progressively diminishes with increase of $^4$He coverage $D_4$,
vanishing at $D_4\simeq 2.5$ layers.  However, such a Curie component
$\chi_C(T)$ is also found in the spin susceptibility of a high-mobility
2D layer in silicon.\cite{rez}  In this case, disorder effects are
suppressed, while the effective interaction between electrons is
strong, since the $r_s$ parameter attains values around 10.
Furthermore, the same behavior of $\chi(T\to 0)$ is
observed\cite{godfrin1,godfrin2} when two monolayers of $^3$He are
absorbed onto a graphite substrate pre-plated with a monolayer of $^4$He.
The density of liquid $^3$He in this experiment is about an order
of magnitude higher than in the work of Ref.~\onlinecite{sheldon}.
Again the effective interaction is strong, since $M^*$ is
drastically enhanced.  Thus, at least two examples exist in
which the interaction appears to be the dominant playmaker,
rather than disorder.

If indeed disorder proves to be irrelevant in some strongly
interacting Fermi systems, how does the interaction give rise
the inferred singular behavior of $\chi(T\to 0)$?  A possible
answer to this question is provided by a microscopic mechanism
proposed over a decade ago,\cite{ks} which predicted a flattening of
the single-particle spectrum in strongly correlated Fermi
systems well before the phenomenon was observed in angle-resolved
photoemission studies.\cite{shen,norm1}  In this scenario, known
as fermion condensation,\cite{vol,noz,physrep,normfc,zkc,ikk,kz}
the flattening effect is associated with a phase transition
in which the conventional Landau state suffers a rearrangement.
It is intrinsic to this model that the Landau-Migdal quasiparticle
picture retains its validity beyond the critical point.  However,
at $T=0$ the Fermi step $n_F(p)=\theta(p_F-p)$ is replaced by a
new momentum distribution, $n_0(p)$, determined by the variational
condition
\beq
\delta E/\delta n(p)=\mu \   .
\label{var}
\eeq
The distribution $n_0(p)$ differs from $n_F(p)$ in a finite momentum
interval $p_i<p<p_f$ that includes the Fermi momentum $p_F$.  Since
the variational derivative $\delta E/\delta n(p)$ is by definition
the energy $\epsilon(p)$ of a Landau quasiparticle, the condition
(\ref{var}) implies that in this domain, the single-particle
spectrum must be completely flat, pinned to the chemical potential
$\mu$:
\beq
\epsilon(p)=\mu \  , \quad p_i<p<p_f \  .
\label{var2}
\eeq
The set of single-particle states having $\varepsilon(p)=\mu$ is
called the fermion condensate (FC) in analogy with the Bose condensate
existing in liquid $^4$He, since both condensates entail similar
$\delta$-like singular contributions to the density of states.

At finite $T$, the degeneracy of the FC spectrum is lifted in
accordance with the formula\cite{noz}
\beq
\xi(p,T{\to}0)\equiv \epsilon(p){-}\mu=T\ln {1{-}n_0(p)\over n_0(p)} \  ,
\quad p_i{<}p{<}p_f \  ,
\label{spect}
\eeq
which ensures consistency between the Fermi-Dirac expression
$n(p,T)=[1+\exp(\xi(p)/T)]^{-1}$ and the solution $n_0(p)$
of Eq.~(\ref{var}) at $T=0$. At higher $T$ the FC density falls to
disappear at the critical temperature $T_f$.

Having set the stage, let us now evaluate the spin susceptibility
$\chi$ of a strongly correlated Fermi system that experiences
a rearrangement of the Landau state at critical temperature $T_f$.
It is instructive to begin on the ``metallic''
side of the transition, where standard Landau theory still
holds.  Within the quasiparticle picture,\cite{lanl}
\beq
\chi = {\chi_0\over 1-g_0\Pi_0} \  ,
\label{hi}
\eeq
where $g_0$ is the zeroth harmonic of the Landau spin-spin interaction
and $\chi_0=-\mu^2_B\Pi_0$ is the spin susceptibility of the noninteracting
quasiparticles in terms of the Bohr magneton $\mu_B$.
The corresponding polarization operator is
\beq
\Pi_0(T)= \int {dn(\xi(p))\over d\xi(p)}
d\tau\equiv v_F^0  P_L I(T) \  ,
\label{pio}
\eeq
where $v_F^0=\partial\varepsilon^0_p/\partial p$ is the quasiparticle
group velocity and $P_L$ is the conventional $T$-independent,
$D$-dimensional Landau result for $\Pi_0(T)$, while
\beq
I(T)= -\int\limits_{-\infty}^{\infty}{dn(\xi, T)\over d\xi}
{dp(\xi)\over d\xi} d\xi  \  .
\label{it}
\eeq
The derivation of Eq.~(\ref{hi}) takes account of the fact that
the renormalization factor $z$ specifying the quasiparticle weight
enters only in the phenomenological parameter $g_0$.  According to
Migdal's relation ${\cal T}({\bm \sigma};k=0,\omega\to 0)=\partial
G^{-1}(\varepsilon)/\partial \varepsilon\equiv z^{-1}$ stemming
from spin conservation, effects of renomalization of the vertex parts
${\cal T}$ and the Green functions $G$ cancel.

Noting that $dn(\xi,T)/d\xi$ decays rapidly at $\xi\geq T$,
Rolle's theorem may be invoked to reduce integral (\ref{it})
to the value of the function $dp(\xi)/d\xi$ at a certain
point $\xi_R\sim T$.
In strongly correlated Fermi systems that exhibit non-Fermi-liquid
behavior, there are two distinct temperature-density regions
in which the function $\left(dp(\xi)/d\xi\right)_{\xi_R}$ shows
qualitatively different departures from the Landau norm.  The first
region is situated close to the critical density $\rho_{\infty}$.
A portion of spectrum $\xi(p)$ adjacent to the Fermi surface is
flattened anomalously, and the curve $dp(\xi)/d\xi$ bends downward,
producing $T$-dependent corrections to $I(T)$, and hence to $\chi(T)$,
that become important at sufficiently high $T$.
In 2D liquid $^3$He, this behavior is successfully described by the
phenomenological formula\cite{dyug}
$\chi(T)\sim \left[(T^{**}(\rho))^2+T^2\right]^{-1/2}$.
Still within the first region, Landau theory fails at low $T \sim 0$
as well.  New terms in the Taylor expansion,
\beq
\xi(p,\rho)=p_F{p-p_F\over M^*(\rho)}+\xi_3{(p-p_F)^3\over p_F^3}
+ \cdots \ ,
\label{exp}
\eeq
odd in $p-p_F$, come into play since the first term is suppressed by
the divergence of the effective mass.  Neglecting the $M^*$ term, the
cubic term becomes dominant and upon inserting Eq.~(\ref{exp})
into Eq.~(\ref{pio}) one finds
$\Pi_0(T\to 0)\sim-T^{-2/3} $. In the second region, $T\ll T_f(\rho)$ and
the FC has come to occupy
a significant fraction of the Fermi sphere.  In this case,
substituting Eq.~(\ref{spect}) into Eq.~(\ref{pio}) yields
$\Pi_0(T\to 0)\sim -1/T $.

\begin{figure}[t]
\includegraphics[width=0.8\linewidth,height=0.7\linewidth]{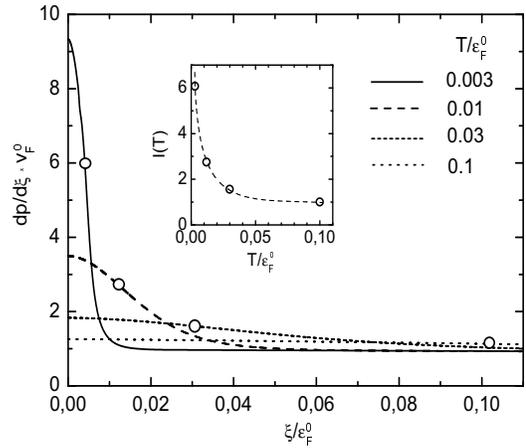}
\caption{Derivative $dp(\xi)/d\xi$, in units of $(v_F^0)^{-1}$, plotted
as a function of $\xi/\varepsilon_F^0$, based on calculations
in the Nozi\`eres model at different temperatures $T$.  Open
circles label the points ($\xi_R$, $dp/d\xi|_R$) on the
corresponding curves.  Inset: The integral $I(T)$, evaluated
as $(dp/d\xi)_{\xi_R}$, is drawn versus $T/\varepsilon_F^0$.
}
\label{fig:fig1}
\end{figure}

The foregoing inferences are confirmed in numerical calculations.
For example, Fig.~\ref{fig:fig1} shows results for the functions
$dp/d\xi$ and $I(T)$ based on the Nozi\`eres model,\cite{noz} in
which the effective interaction between particles is proportional to
$\delta({\bf p}_1-{\bf p}_2)$.  The transitional behavior near
$\rho_\infty$ is absent in this model, and the function $I(T)$
is simply the sum of a term $\sim 1/T$ and a constant term.

In arriving at these results, we have neglected damping, which is
more pronounced in systems containing a FC than in ordinary
Fermi liquids.  However, in the energy region $\varepsilon\simeq T$
relevant to the current problem, the damping parameter is moderate,
with $\gamma(T)\sim T$.  As a consequence, damping does not affect
the predicted form of $\Pi_0(T\to 0)$, instead altering only
numerical factors.\cite{schuck,kz}  In the transitional
region where $\xi(p)$ is not completely flat, the ratio
$\gamma(T)/T$ is below 1; hence the inclusion of damping has
little effect.

We conclude that there exist two regimes in which the spin susceptibility
shows characteristic variation with temperature, in addition
to the standard Fermi-liquid domain.  The first is a transitional
region close to the point of fermion condensation, where the FC is
incipient or still minute.  Upon inserting the above result
$\Pi_0(T\to 0)\sim -T^{-2/3}$ into the Eq.~(\ref{hi}), simple
algebra leads to
\beq
\chi(T\to 0)\sim (C_{\rm t}T^{2/3}-\Theta_{\rm t})^{-1} \  ,
\label{hitr}
\eeq
where $\Theta_t\sim -g_0\rho$.
Such a power-law $T$-dependence is observed in
a number of heavy-fermion metals.\cite{coleman,gegenwart}

As the density increases, we pass from the transitional regime
characterized by (\ref{hitr}) to the region where the FC contribution
plays the dominant role.  On combining Eqs.~(\ref{hi}),
(\ref{pio}), and the result $\Pi_0(T \to 0) \sim -1/T$,
we obtain the Curie-Weiss formula
\beq
 \chi(T\to 0)\sim \left(T-\Theta_W \right)^{-1} \  ,
\label{hifc}
\eeq
with a Weiss temperature $\Theta_W\sim-g_0(\rho)\rho$.
Remarkably, working within the general framework of the Landau
quasiparticle approach, we have found conditions under which a
strongly correlated Fermi liquid behaves as a gas of localized spins.

\begin{figure}[t]
\includegraphics[width=0.8\linewidth,height=0.6\linewidth]{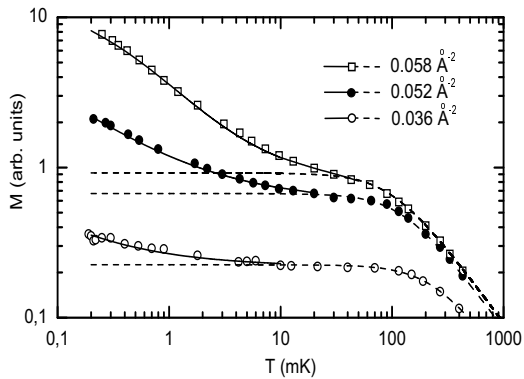}
\caption{Magnetization for $^3$He films at different densities.
Experimental data from Refs.~\onlinecite{godfrin1,godfrin2} are
indicated by symbols, solid curves show the present predictions at
low $T$, and dashed lines indicate the phenomenological fit
of Ref.~\onlinecite{dyug} at higher $T$.
}
\label{fig:fig2}
\end{figure}

As a rule, the value of the Weiss temperature $\Theta_W$ is small
(e.g., $\Theta_W\simeq -0.3$\,mK for the uppermost curve in
Fig.~\ref{fig:fig2}).
The reason for this is that there is no ferromagnetic instability
in any of the systems under consideration, i.e.\ the Pomeranchuck
stability condition $1+g_0N(0)>0$ is not violated. Since the density of
states $N(0)$ at the Fermi surface is proportional to $M^*$,
preservation of this condition implies that $g_0(\rho)$ changes
its sign at the critical density $\rho_{\infty}$ where $M^*\to \infty$.
Hence, at this density the character of the correlations changes from
ferromagnetic to antiferromagnetic; such a change does take place
in strongly correlated 2D liquid $^3$He (see Fig.~1 of
Ref.~\onlinecite{godfrin1}).

The predicted behaviors (\ref{hitr}) and (\ref{hifc}) explain the
measurements  that trace the gradual evolution of the spin
susceptibility of a 2D electron layer in silicon as the density
$\rho$ is lowered.\cite{rez}  The constant susceptibility
of the ordinary Fermi liquid gives way to the $T$-dependent behavior
$\chi(T)\sim T^{-1}$ of a gas of localized spins.
The same trends are seen in the extensive data on the magnetization
of $^3$He films reported in Refs.~\onlinecite{godfrin1,godfrin2}, some
of which are plotted in Fig.~\ref{fig:fig2} along with our
theoretical predictions.

Before focusing on these results, it must be remarked that
comparison of different sets of data on 2D liquid $^3$He
reveals a crucial dependence of the effective-mass enhancement
on the number of $^4$He layers separating the $^3$He monolayer
from the graphite substrate\cite{morishita}: The smaller the number,
the greater is the increase of $M^*$.  It would appear that the
enhancement of $M^*$ is somehow associated with the crystal lattice
of the substrate.  A similar situation occurs in the flattening of
single-particle spectra of 2D superconductors.  For example, in
the case of a quadratic lattice, flattening is most evident in
the region adjacent to the van Hove points.\cite{shen,norm1}
As seen in ARPES data\cite{camp} on Na$_{0.5}$VO$_2$, flat portions
of the single-particle spectrum exist in 2D metals having the same
Brillouin zone as 2D liquid $^3$He.   Therefore we suggest that
the formation of FC in certain segments of the Fermi line
of 2D liquid $^3$He and, correspondingly, the emergence of Curie-Weiss
components in its spin susceptibility, begins at densities markedly
lower than the threshold for fermion condensation in homogeneous
2D liquid $^3$He.

Turning now to Fig.~\ref{fig:fig2}, we first consider the uppermost
curve, belonging to the highest density $\rho=0.058$\,\AA$^{-2}$,
where it is assumed that the FC occupies a significant part of
the Fermi sphere. The corresponding contribution to the spin
susceptibility $\chi(T\to 0)$ is evaluated by means of Eq.~(\ref{hifc}).
To accomodate higher $T\sim T_f$, we suppose that upon approach to
$T_f$ the FC density behaves as $(1-T^2/T^2_f)^{1/2}$,
with $T_f= 30$\,mK.  (In fact, the specific form of this
attenuation factor is immaterial.)  The contribution to $\Pi_0(T)$
from the remaining, normal part of the Fermi liquid is assumed
to be constant at a value $\simeq 0.92$.  Given these two parameter
choices, the experimental data for $\chi(T,\rho=0.058$\,\AA$^{-2})$
are well reproduced over more than two orders of magnitude
on the mK scale.

The other two curves in Fig.~\ref{fig:fig2}, for the densities
$\rho=0.052$\,\AA$^{-2}$ and $0.036$\,\AA$^{-2}$, are determined
under the assumption that FC is poised on the verge of onset,
so that the cubic term prevails in the series representation
(\ref{exp}) of the spectrum $\xi(p)$.
As for the uppermost curve, the effect of flattening
on $\chi(T \to 0)$ in this transition regime is attenuated with
rising $T$ through a factor $(1-T^2/T^2_{\rm t})^{1/2}$.
The constant contribution to $\Pi_0(T)$ from other effects
is taken as $\simeq 0.04$ for $\rho=0.052$\,\AA$^{-2}$,
and $\simeq 0.014$ for $\rho=0.036$\,\AA$^{-2}$.
The remaining parameter choices are:
$T_{\rm t}=20$\,mK, $C_{\rm t}=1.2\,$mK$^{1/3}$ for $\rho=0.052$\,\AA$^{-2}$;
$T_{\rm t}=10$\,mK, $C_{\rm t}=17$\,mK$^{1/3}$ for $\rho=0.036$\,\AA$^{-2}$;
and $\Theta_{\rm t} = 0$ at both densities.

When the temperature is lowered to values such that $T<T_Z=\mu_BB$,
where $B$ is the strength of a magnetic field applied to the sample,
the perturbative approach employed above in evaluating $\chi(T)$
becomes questionable.  One then needs to solve equations for the
spectra $\xi_{\pm}(p)$ directly.  These equations may be written in
the form
\begin{eqnarray}
\xi_+(p)&=&\xi^0_p+ {1 \over 2} T_Z
+\int f({\bf p},{\bf p}_1){1\over 2}\left[n_+(p_1){+}n_-(p_1)
\right]d\tau_1 \  , \nonumber\\
\xi_-(p)&=&\xi_+(p)-T_Z \  ,
\label{spinsy}
\end{eqnarray}
where $n_+$ and $n_-$ are the quasiparticle momentum distributions of the
subsystems with spin projections $\pm {1\over 2}$.  The interaction
function $f$ is assumed to be independent of the momentum distributions
$n_+$ and $n_-$ themselves.

\begin{figure}[t]
\includegraphics[width=0.8\linewidth,height=0.6\linewidth]{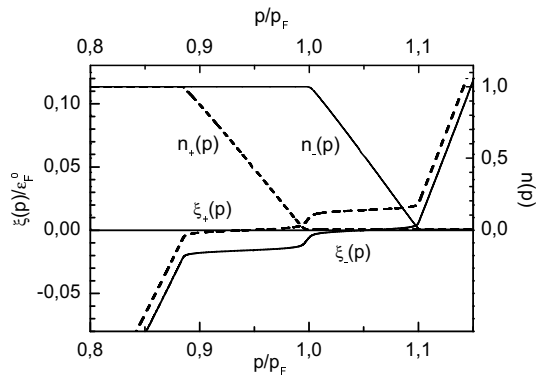}
\caption{Single-particle spectra $\xi_+(p)$ (dashed line) and
$\xi_-(p)$ (solid line) in units of $\varepsilon_F^0$, together
with momentum distributions $n_+(p)$ (dashed line) and $n_-(p)$
(solid line), plotted as functions of $p/p_F$ for
$T_z/\varepsilon_F^0=10^{-2}$ and $T/\varepsilon_F^0=10^{-3}$.
The Nozi\`eres model is assumed.
}
\label{fig:fig3}
\end{figure}

If the field $B$ is small, the solution of Eqs.~(\ref{spinsy})
can be expressed in terms of the solution of the FC problem
\beq
\xi(p)=\xi^0_p + \int f({\bf p},{\bf p}_1)\,n_0(p_1)\,d\tau_1 \
\label{eqsam}
\eeq
in the absence of the field.  Setting $B=0$ in Eqs.~(\ref{spinsy}),
we are led to the conclusion that the sum $\left[n_+(p) + n_-(p)\right]/2$
must coincide with the solution $n_0(p)$ of Eq.~(\ref{eqsam}).
But according to the second of Eqs.~(\ref{spinsy}), only one of
the two states with the same momentum $p$ can belong to the FC.  In
particular, if $n_0(p)>1/2$, then $n_-(p)=1$ and $n_+(p)=2n_0(p)-1$;
hence, in the FC region one obtains $\xi_-(p)\simeq -T_z$,
while $\xi_+(p)\simeq T\ln [(1{-}n_0(p))/(n_0(p){-}1/2)]$.
Conversely, if $n_0(p)<1/2$, then $n_-(p)=2n_0(p)$
and $n_+(p)=0$; hence, in the FC region one has
$\xi_-(p)\simeq T\ln [(1/2{-}n_0(p))/n_0(p)]$, while $\xi_+(p)\simeq T_Z$.
Thus we infer that at the point where $n_0(p)=1/2$,
there is a jump in the single-particle energies.  This behavior is
illustrated in Fig.~\ref{fig:fig3}, where the particle energies
$\xi_+(p)$ and $\xi_-(p)$ are plotted for the Nozi\`eres model
with coupling constant $f/\varepsilon_F^0=0.2$.

Having obtained these results, the FC magnetization ${\cal M}^f$
is readily evaluated as sum of the integral over $n_0(p)$ from
$n_0(p)=0$ to $1/2$, and the integral of $1-n_0(p)$ from
$n_0(p)=1/2$ to $1$.  Thus we see that in the case $\mu_B B>T_Z$,
the FC magnetization ${\cal M}^f$ is independent of both $T$ and $B$.
This property implies that the spin susceptibility, determined as
the derivative $\chi=\partial {\cal M}/\partial B$, does not contain
the FC contribution.  We therefore conclude that at finite values of
$B$ and low $T$, the spin susceptibility $\chi(T)$ of a system
containing a FC increases with temperature until $T$ attains
values comparable to $T_Z$, and then begins to fall off as
$T^{-1}$ if the FC fraction is significant.

In summary, we have shown that if the interaction between particles
is strong enough to induce a rearrangement of the Landau state, then
the behavior of the spin susceptibility $\chi(T \to 0)$ progressively
changes from that inherent in Landau Fermi-liquid theory to the
behavior characteristic of a gas of localized spins.  This finding
calls for reexamination of many aspects of the theory of metallic
oxides and heavy metals with localized spins.

We have benefited from valuable discussions with P.~W. Anderson,
L.~P.~Gor'kov, G.~Kotliar, E.~Krotscheck, M.~R.~Norman, and
V.~M.~Yakovenko.  This research was supported by NSF Grant
PHY-0140316 (JWC and VAK), by the McDonnell Center for the
Space Sciences (VAK), and by the Grant NS-1885.2003.2 from the
Russian Ministry of Industry and Science (VAK and MVZ).

\end{document}